# Study of gold induced heavy ion collisions using isospin dependent QMD model

Bhawna Sharma[1],[*] Sanjeev Kumar[1], Suneel Kumar[1] and Rajeev K. Puri[2]
[1]*School of Physics and Materials Science, Thapar University, Patiala – 147004 (Punjab) INDIA and*
[2]*Department of Physics, Panjab University, Chandigarh – 160014 INDIA*
* email:bhawna.dutt87@gmail.com

## Introduction:

One of the main interests to study the heavy ion collision at intermediate energy is to study the nuclear matter at extreme conditions and excitation energy. When two nuclei collide, formation of compound nucleus takes place then this compound nucleus breaks into several medium, light mass fragments along with the emission of free nucleons in all possible directions. This phenomenon is termed as 'Multifragmentation'. The observation of multifragment configuration would correspond to the state of matter intermediate between nuclear liquid (nucleus close to its ground state) and nuclear vapor (assembly of nucleons and lighter mass fragments at high temperature). The hope to establish a link to the liquid-gas phase transition in the nuclear matter has been the major motivation for the research and study of multifragment decay of the heavy nuclei in recent years. The multifragmentation studied by experimentalist vastly in the past decade for symmetrical and asymmetrical systems [1]. The findings of the symmetrical systems are verified by theoretician many times, but, limited efforts were done to verify the data of asymmetric colliding nuclei.

We have studied the fragment production mechanism in the set of four reactions $^{197}Au_{79}+^{12}C_6$, $^{197}Au_{79}+^{26}Al_{13}$, $^{197}Au_{79}+^{63}Cu_{29}$ and $^{197}Au_{79}+^{208}Pb_{82}$. The reactions are simulated at an energy 600 MeV/nucleon and collision geometry is varied from central to peripheral ($\hat{b}$ = b/b$_{max}$ = 0 to 1). A theoretical investigation has been carried out on the study of mass dependence of intermediate mass fragments ($5 \leq A \leq A_{tot}/6$) and other fragments. This work has been carried out within the frame work of isospin dependent quantum molecular dynamics (IQMD) model.

## The Model: IQMD

The IQMD model has been successfully used for the analysis of large number of observables. It is a N-body theory which simulates heavy ion reaction at intermediate energies on an event by event basis [2]. This is an improved version of QMD model and is based on molecular dynamics picture & it includes
1. Isospin dependent Coulomb potential
2. Symmetry Potential
3. N-N cross-sections

An IQMD simulation needs three steps which are explained below

1) Initialization of nucleon: In this step, each nucleon is represented by Gaussian wave packet given as:

$$f_i(\vec{r},\vec{p},t) = \frac{1}{(\pi\hbar)^3} \times e^{[-(\vec{r}-\vec{r}_i(t))^2 \frac{2}{L}]} \times e^{[-(\vec{p}-\vec{p}_i(t))^2 \frac{L}{2\hbar^2}]}$$

2) Propagation

The nucleons are propagated under the total interaction calculated by the Hamiltonian equations of motion:

$$\frac{dr_i}{dt} = \frac{d\langle H\rangle}{dp_i}, \quad \frac{dp_i}{dt} = -\frac{d\langle H\rangle}{dr_i}$$

Where $\langle H \rangle = \langle V \rangle + \langle T \rangle$ is the Hamiltonian $V^{ij} = V^{ij}_{Skyrme}+V^{ij}_{Yukawa}+V^{ij}_{mdi}+V^{ij}_{Coul}+V^{ij}_{sym}$ is the total interaction potential.

3) N-N Collision

Collision between two nucleons takes place only if

$$|\vec{r}_i - \vec{r}_j| \leq \sqrt{\frac{\sigma_{tot}}{\pi}}$$

## Results and discussion:

In the fig.1, we have displayed the multiplicity of free nucleons and light mass fragments (LMF's) as a function of the system mass at the different impact parameters for three ($^{197}Au_{79}+^{12}C_6$, $^{197}Au_{79}+^{26}Al_{13}$ and $^{197}Au_{79}+^{63}Cu_{29}$) different reactions. Both free nucleons and LMF's show increasing trends for central collisions. With the increase in the size of system, number of the participant nucleons increases. This will lead to more thermalization of the system. Due to this reason, increase in multiplicity of these fragments which originate from the participant zone. But as one move towards the peripheral collisions, the amount of free nucleons and LMF's decreases as participant region decreases.

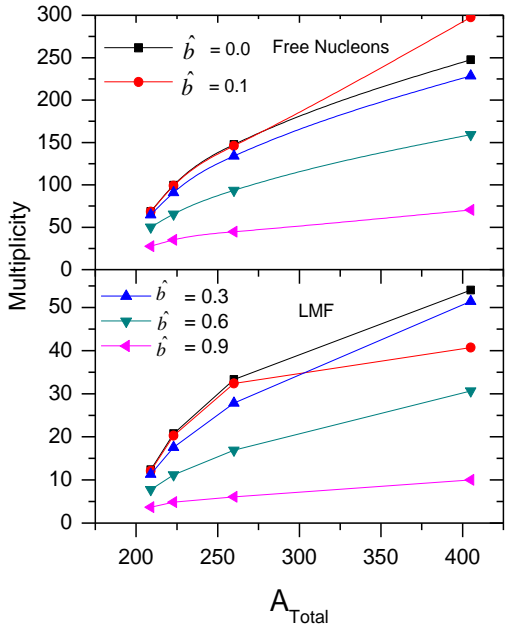

**Fig. 1** Multiplicity of free nucleons and LMF's ($2 \leq A \leq 4$) as a function of system mass at different scaled impact parameters

The multiplicity of IMF's as a function of $Z_{bound}$ for targets Cu and Pb is displayed in figure 2. The quantity $Z_{bound}$ is defined as sum of all atomic numbers $Z_i$ of all projectile fragments with $Z_i \geq 2$. It is observed that multiplicity shows a good agreement for low $Z_{bound}$, but it fails for high $Z_{bound}$. This failure is due the method of analysis MST which we had used in our analysis, because MST method gives one heavy cluster at the time of high density. The discrepancy between theory and experiments can be removed by using reduced isospin dependent NN cross section and sophisticated clustrization algorithm SACA [3].

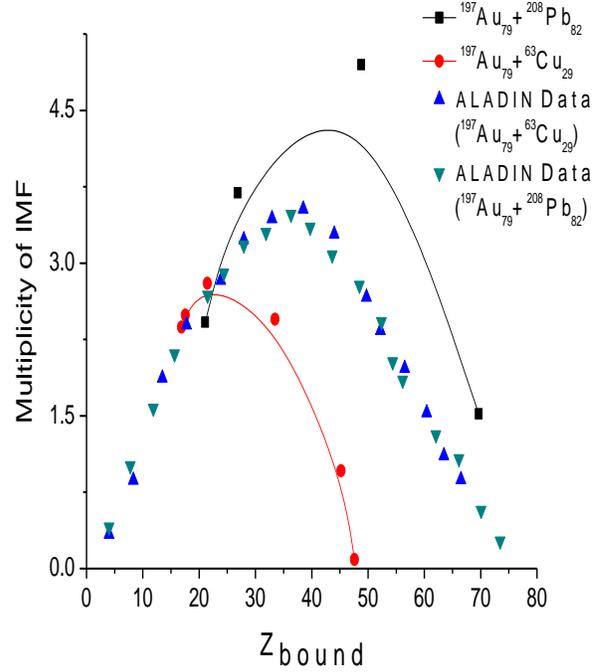

**Fig. 2** Multiplicity of IMF's as a function of $Z_{bound}$

## References

[1] C. A. Ogilvie et al., Phys. Rev. Lett. **67**, 1214 (1991).
[2] Ch. Hartnack, Rajiv K Puri, J Aichelin, J. Konopka, S.A.Bass, H.Stocker, W. Greiner, Eur Phys. J. A 1 **151** 169 (1998).
[3] R.K. Pur and J. Aichelin, J. Comp. Phys. 162, 245 (2000).